\documentclass[a4paper,11pt]{article}
\usepackage{pos}
\usepackage{xspace}
\usepackage{hyperref}
\usepackage{mathptmx}

\usepackage[left]{lineno}
\newcommand{\sib}[1]{\textsc{Sibyll}\,#1\xspace}
\newcommand{\qgsii}{\textsc{QGSJet~II-04}\xspace}
\newcommand{\eposlhc}{\textsc{Epos-LHC}\xspace}
\newcommand{\Xmax}{\ensuremath{X_\text{max}}\xspace}
\newcommand{\EeV}{\,\text{EeV}\xspace}
\newcommand{\eV}{\,\text{eV}\xspace}
\newcommand{\gcm}{\,\text{g}/\text{cm}^{2}}
\newcommand{\avg}[1]{\ensuremath{\langle{#1}\rangle}}
\newcommand{\DeltaXmax}{\ensuremath{\Delta\Xmax}\xspace}
\newcommand{\Rhad}{\ensuremath{R_\text{had}}\xspace}

\title{Overview of hadronic interaction studies at the Pierre Auger Observatory}
\author{Jakub Vícha$^{a,*}$ for the Pierre Auger Collaboration$^{b}$}

\affiliation[a]{FZU - Institute of Physics of the Czech Academy of Sciences, Na Slovance 1999/2, 182 00, Prague 8, Czech Republic}
\affiliation[b]{Observatorio Pierre Auger, Av. San Martín Norte 304, 5613 Malargüe, Argentina\\
Full author list: \href{https://www.auger.org/archive/authors_2024_11.html}{https://www.auger.org/archive/authors\_2024\_11.html}
}

\emailAdd{spokespersons@auger.org}

\abstract{The combination of fluorescence and surface detectors at the Pierre Auger Observatory offers unprecedented precision in testing models of hadronic interactions at center-of-mass energies around 70 TeV and beyond. However, for some time, discrepancies between model predictions and measured air-shower data have complicated efforts to accurately determine the mass composition of ultra-high-energy cosmic rays. A key inconsistency is the deficit of simulated signals compared to those measured with the surface detectors, typically interpreted as a deficit in the muon signal generated by the hadronic component of simulated showers.

Recently, a new global method has been applied to the combined data from the surface and fluorescence detectors at the Pierre Auger Observatory. This method simultaneously determines the mass composition of cosmic rays and evaluates variations in the simulated depth of the shower maximum and hadronic signals on the ground. The findings reveal not only the alleviated muon problem but also show that all current models of hadronic interactions predict depths of the shower maximum that are too shallow, offering new insights into deficiencies in these models from a broader perspective.}

\FullConference{7th International Symposium on Ultra High Energy Cosmic Rays (UHECR2024)\newline
 17-21 November 2024\newline
Malargüe, Mendoza, Argentina}

\begin{document}
\maketitle

\section{Introduction}
The origin of ultra-high-energy cosmic rays (UHECR, $>10^{18}\eV$), especially at the highest energies ($>10^{20}\eV$), is still a mystery. 
The key knowledge to reveal this puzzle is their mass composition needed to account for propagation effects and identify the directions of sources in the arrival directions after accounting for bending effects in the magnetic fields.
So far, no obvious clustering of events above 100~EeV has been observed \cite{ToshiICRC2023} and no obvious source candidates are associated with arrival directions of UHECR, even when the effect of the Galactic magnetic field is accounted for the most energetic one \cite{AmaterasuBacktracking}.
The cosmic-ray mass is also crucial to assess the acceleration capabilities of source candidates.

The mass of UHECR can be estimated from the comparison of measured properties of air showers, like the depth of shower maximum (\Xmax) and the number of muons on ground, with the predictions of models of hadronic interactions.
The main problem of this interpretation is in the extrapolation of hadronic interactions that are studied at man-made accelerators at much lower energies and different kinematic regions.

The Pierre Auger Observatory \cite{PACosmicObservatory} can measure the muonic content directly using the underground muon detectors or using the water-Cherenkov stations for highly-inclined air showers (zenith angle $\theta>62^\circ$) when the electromagnetic part of the air shower is heavily suppressed for the time when it reaches the ground.
The measurement of the longitudinal profile using fluorescence telescopes provides a precise estimate of \Xmax.
This detection technique, however, provides smaller event statistics due to the limited duty cycle ($\sim15\%$).

In this proceedings, we focus on the main analyses showing a tension with predictions of models of hadronic interactions.
We start with the discrepancies observed in the muon size, followed by inconsistencies seen in the interpretation of measured \Xmax moments.
We describe the recently published method studying model predictions from the muon size and \Xmax simultaneously in the last part.

\section{Inconsistencies between Models and Measurements of Muons}
\begin{figure}
    \includegraphics[width=0.5\textwidth]{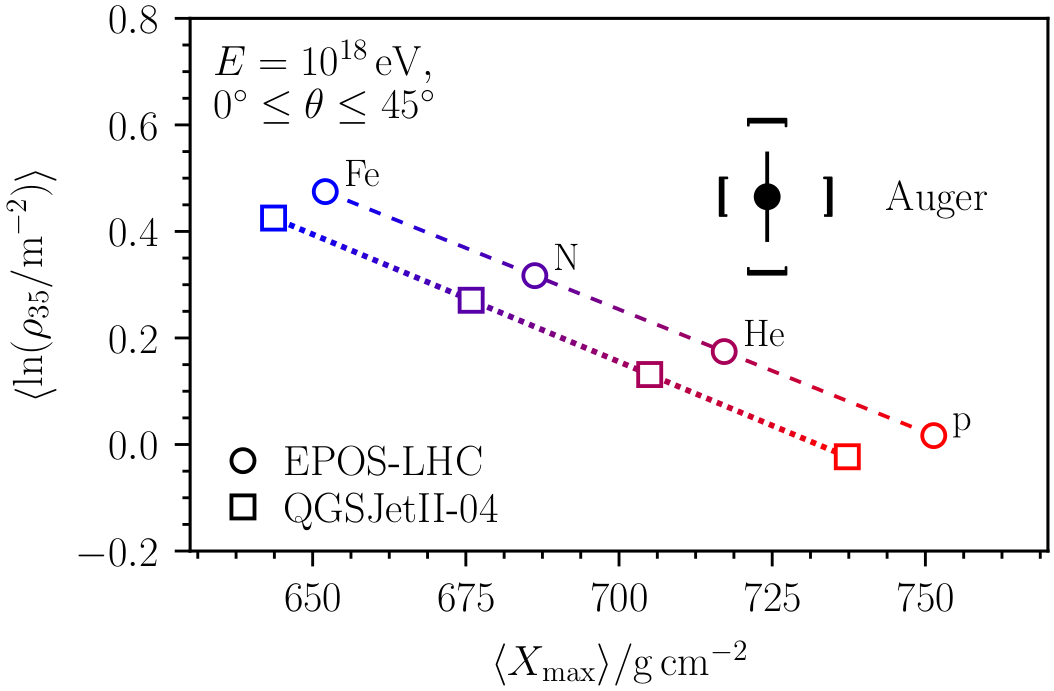}\hspace{0.2cm}
    \includegraphics[width=0.48\textwidth]{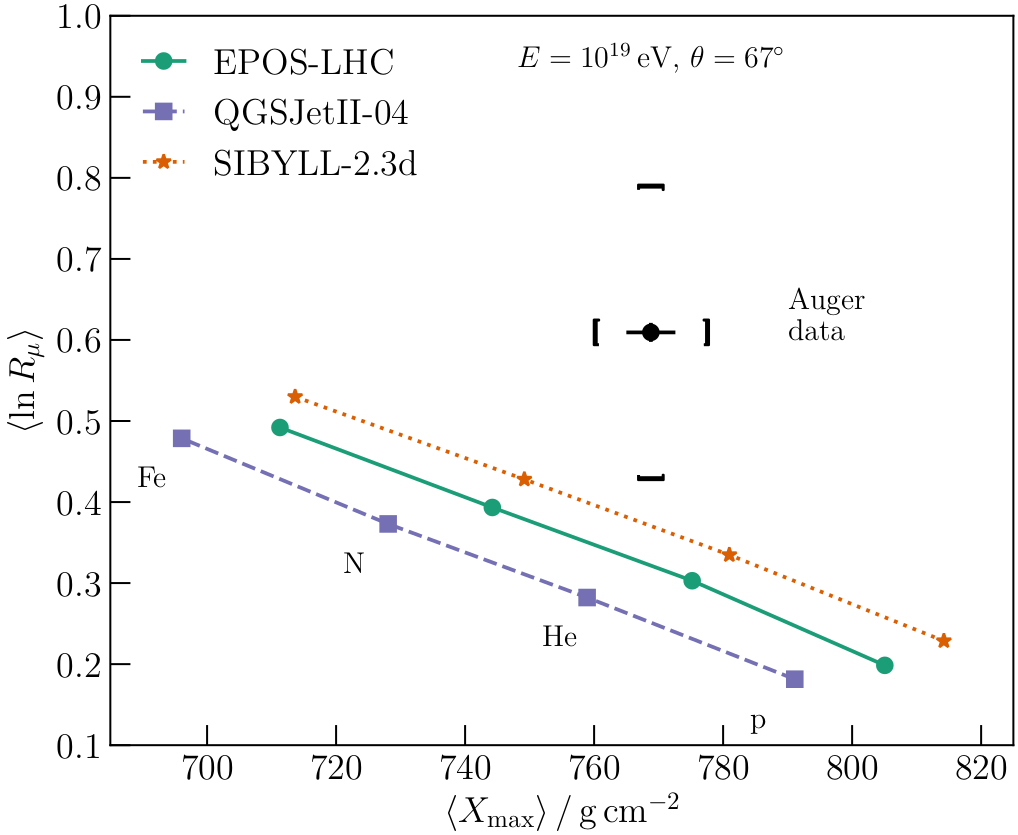}\\ [0.2cm]
    \caption{Comparison of direct measurements of the muon size at the Pierre Auger Observatory with predictions of models of hadronic interactions for combinations of four primary species at given \avg{\Xmax}. The measurements using buried scintillators at $10^{18}\eV$ within 45$^\circ$ of zenith angle (left) and using highly-inclined showers above 62$^\circ$ at $10^{19}\eV$ (right) are shown. Plots are coming from \cite{AmigaMuons} and \cite{MuonFluct2020}.}
    \label{fig:Muons}
  \end{figure}

Inconsistencies in the mass-interpretations of the size of the muon signal from underground detectors \cite{AmigaMuons} and from inclined showers \cite{MuonFluct2020} with respect to \Xmax have been observed, being at a level of $<3\sigma$, see Fig.~\ref{fig:Muons}.
%There are $<3\sigma$ inconsistencies observed in the mass-interpretations of the size of the muon signal measured in the muon detectors below the ground \cite{AmigaMuons} and for inclined showers measured by the surface detectors \cite{MuonFluct2020} with respect to the expectations from the \Xmax measurements, see Fig.~\ref{fig:Muons}.
These inconsistencies have been measured at various energies and zenith angles, and also in surface-detector signals at lower zenith angles where the total signal is induced by a combination of muonic and electromagnetic particles \cite{TestingHadronicInteractions}.
Given the \Xmax expectations, the size of the muon component predicted by the models of hadronic interactions needs to be increased by around 30-60\%.
Note that there is about $30\gcm$ difference in the \avg{\Xmax} predicted by different models of hadronic interactions, which indicates that the predicted \Xmax scale might be also another source of these observed inconsistencies.

On the other hand, the measured fluctuations of the number of muons at $10^{19}\eV$, see Ref.~\cite{MuonFluct2020}, are consistent with expectations for the mass composition obtained from the \Xmax fits, which indicates that the muon problem in air-shower modelling is connected rather to cumulative effects in a series of hadronic interactions than to properties of the first interaction only.

%There are also other indications that the modelling of ground-signal is not corresponding to the measurements. 

\section{Inconsistencies between Models and Measurements of \Xmax}
\begin{figure}
    \includegraphics[width=1.0\textwidth]{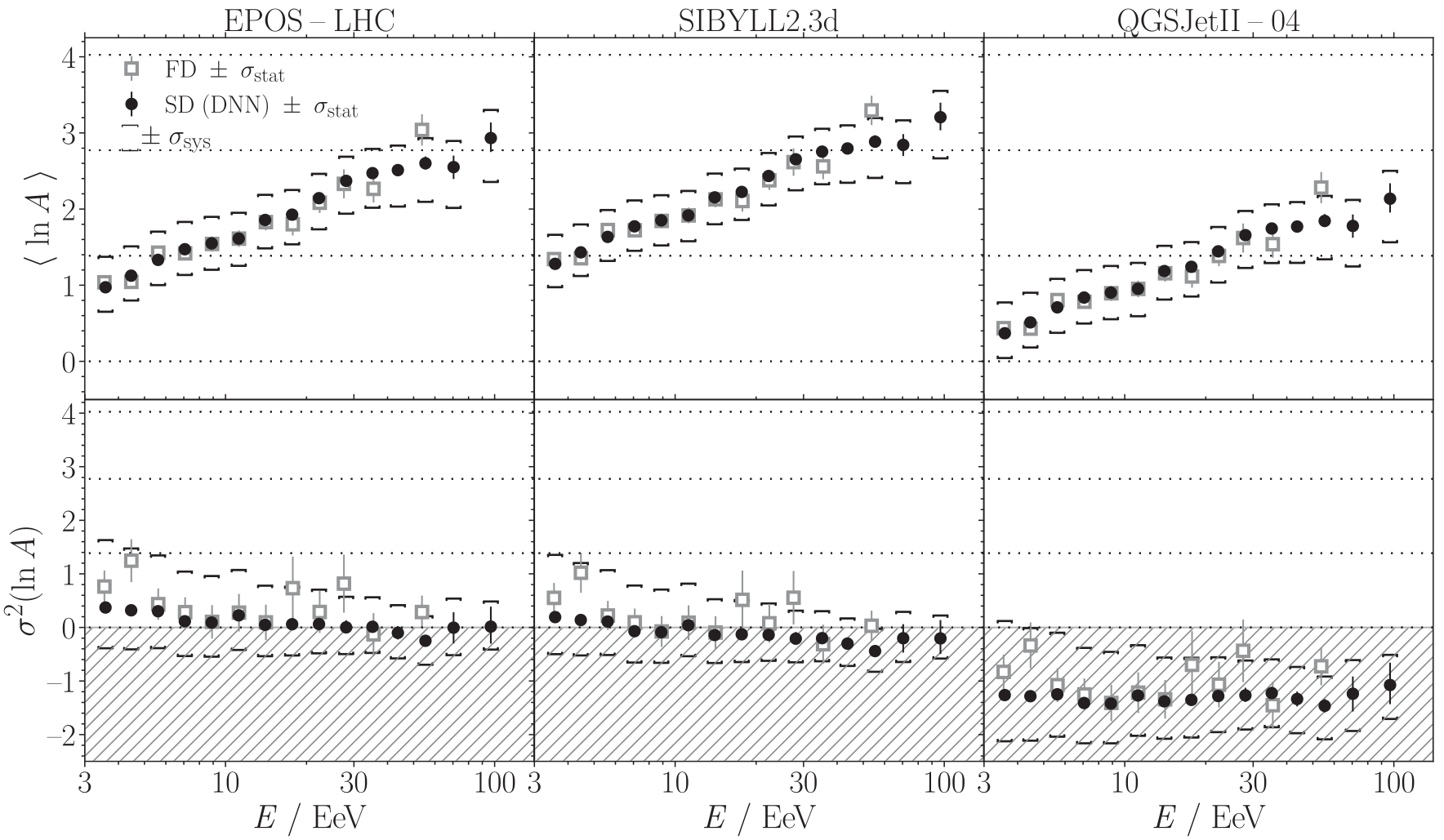}
    \caption{The mean (top) and variance (bottom) of the logarithmic mass number interpreted from the measurements of \avg{\Xmax} and $\sigma\left(\Xmax\right)$ using fluorescence (FD) and surface (SD) detectors is plotted as a function of the reconstructed shower energy for three models of hadronic interactions (from left to right). The plot is taken from \cite{AugerDNN_PRD2025}.}
    \label{fig:LnAmoments}
  \end{figure}

Recently, an unprecedented improvement in the event statistics has been achieved by applying Deep Neural Network (DNN) to estimate the first two \Xmax moments from the recorded surface-detector traces \cite{AugerDNN_PRD2025}.
The interpreted variance of $\ln A$ was found negative (unphysical) in case of the \qgsii model, see Fig.~\ref{fig:LnAmoments}.
The $\ln A$ fluctuations expected from a composition derived from the \avg{\Xmax} measurement are therefore in tension for this model.
In case of the models \eposlhc and \sib{2.3d}, $\sigma^{2}\left(\ln A\right)\simeq0$ is compatible with a pure beam of nuclei above $\sim6\EeV$, however with some possible increase given the systematic uncertainty.
In the same energy region, a gradual increase of \avg{\ln A} is manifested, approximately, from predictions for pure helium ($\ln A \approx 1.39$) to pure nitrogen nuclei ($\ln A \approx 2.64$).
The model predictions of these two models are at the edge of the systematic uncertainty range of the data.
%It is highly unphysical to observe a gradual change from a beam of pure helium to pure lithium, pure beryllium nuclei etc, as pointed out in Ref.~\cite{HeavyMetal_UHECR2024}.
%The $\ln A$ fluctuations expected from a composition derived from the \avg{\Xmax} measurement are therefore in tension with the model predictions of \eposlhc and \sib{2.3d} as well, unless the systematic uncertainty on $\sigma^{2}\left(\ln A\right)$ is considered.

%A part of this mismatch could originate from differences in the \Xmax scale in measured data and simulations. 
Too small values of $\sigma^{2}\left(\ln A\right)$, such as very prominently visible for the \qgsii model, can originate from an incorrect \Xmax scale in simulations.
An increase of the interpreted $\sigma^{2}\left(\ln A\right)$, which would correspond to a more realistic scenario, can be obtained if the \Xmax scale of model predictions gets deeper (as a consequence of the transformations between the \Xmax and $\ln A$ moments in Ref.~\cite{InterpretationOfXmax}).

  \begin{figure}
  \begin{center}
      
    \includegraphics[width=0.45\textwidth]{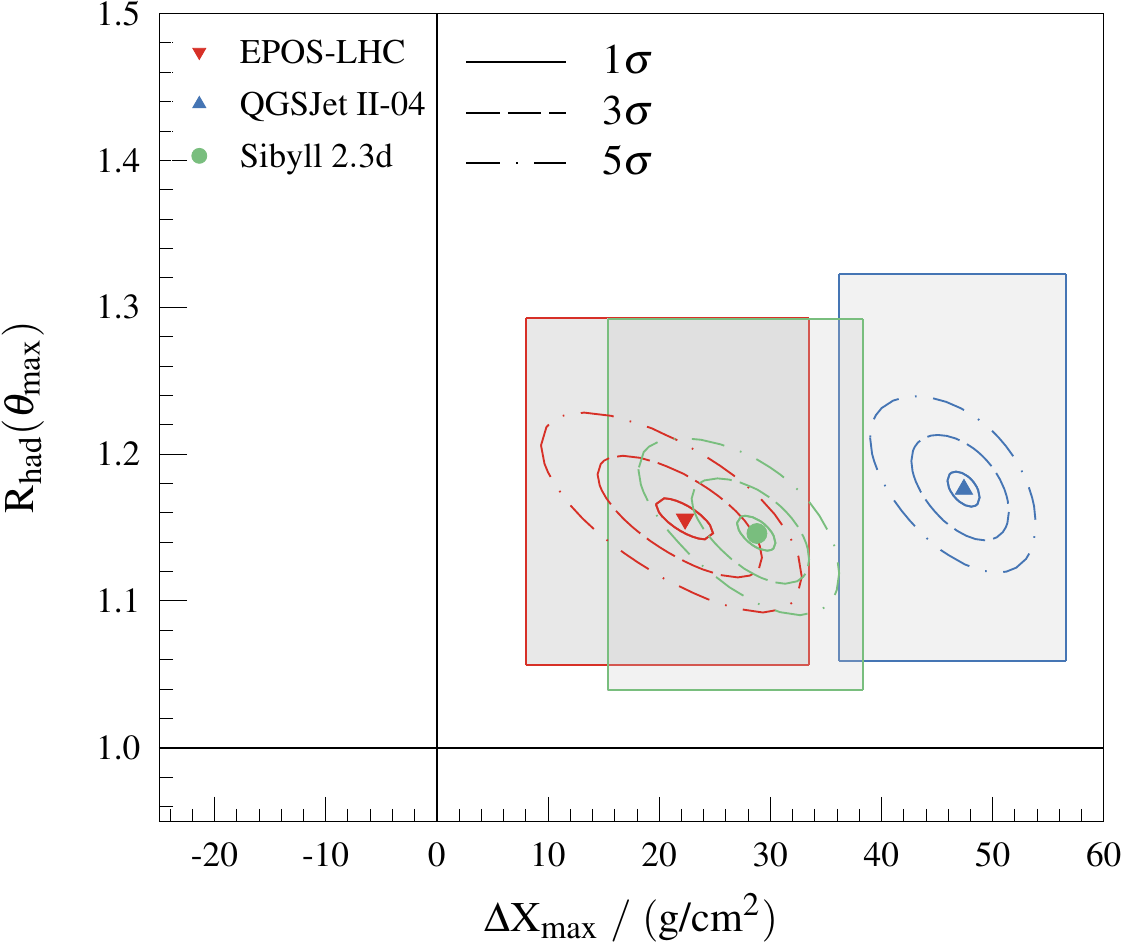}\hspace{0.2cm}
    \includegraphics[width=0.45\textwidth]{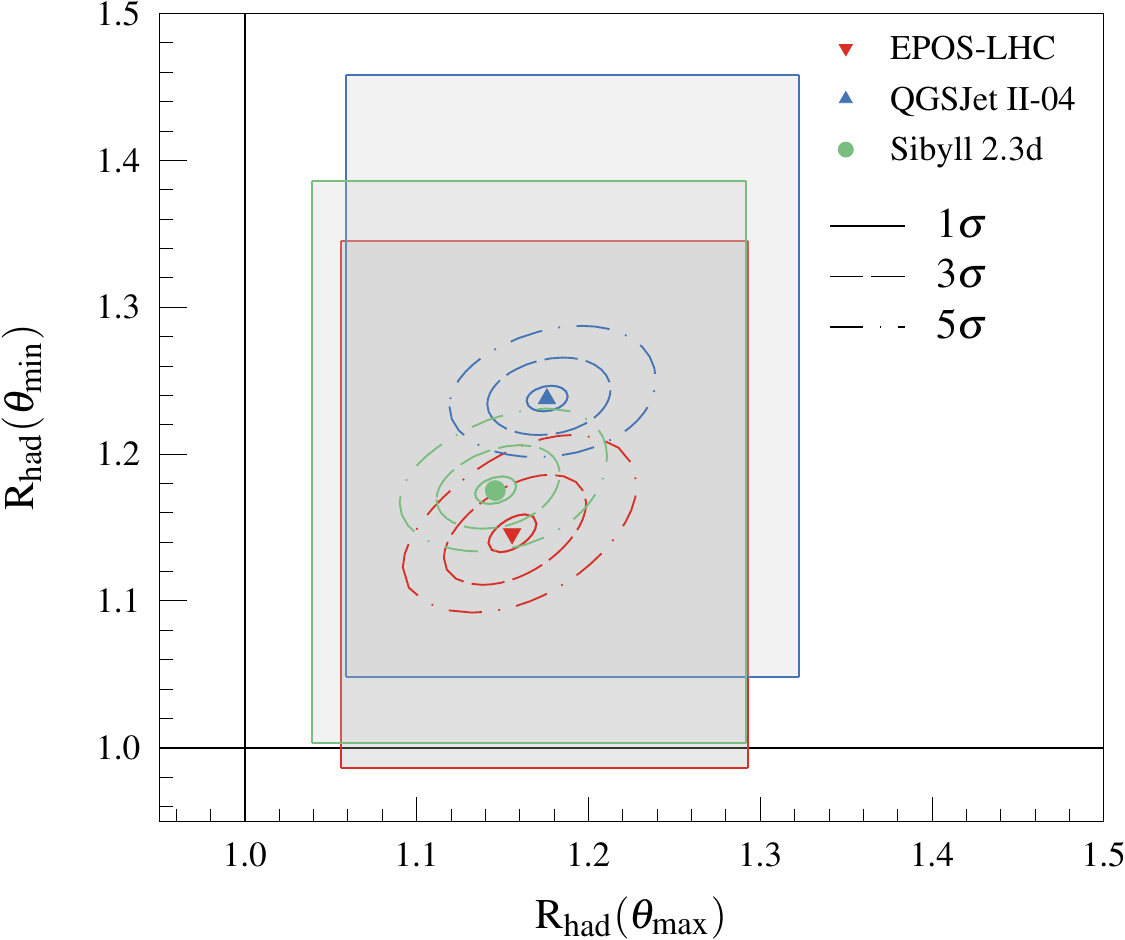}\\ [0.2cm]
    \includegraphics[width=0.9\textwidth]{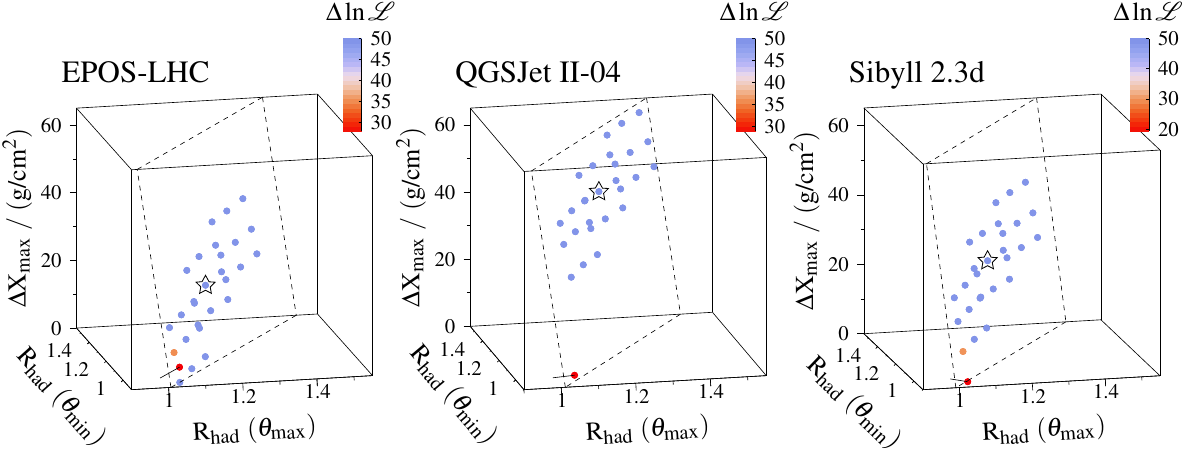}\\ [0.5cm]
    \includegraphics[width=0.9\textwidth]{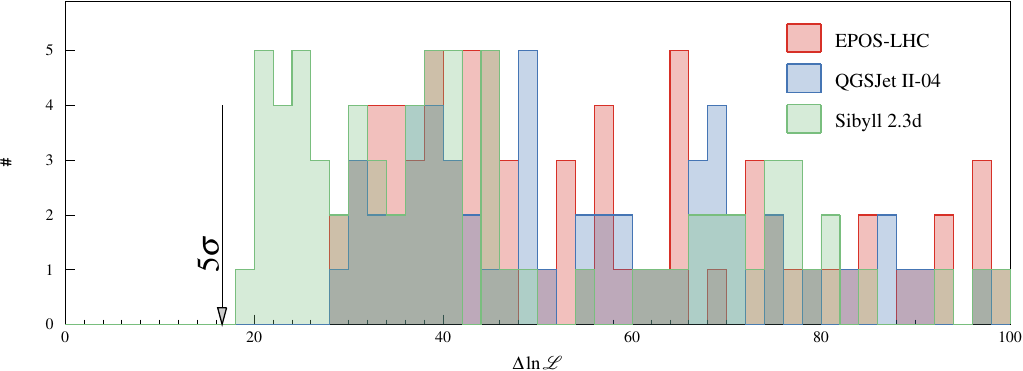}
  \end{center}
    \caption{Top panels: The relations between the modification parameters of the hadronic part of ground signal at the two extreme zenith angles $R_\text{had}\left( \theta_\text{max}{\approx}55^\circ \right)$, $\Rhad\left(\theta_\text{min}{\approx}28^\circ\right)$ and the shift of \Xmax applied to the predictions of the three models of hadronic interactions (\DeltaXmax). % after being fitted to the two-dimensional distributions of measured ground signal and \Xmax.
    Middle panel: The modification parameters for the three models fitted after applying all the combinations of experimental systematic shifts to the measured energy ($\pm14\%$), ground signal ($\pm5\%$) and \Xmax ($^{\text{+}8\gcm}_{-9\gcm}$). All these points lie approximately in a plane (cuts of planes indicated by dashed lines). The color of the points corresponds to the change of the log-likehood expression with respect to the fits with unmodified model predictions. Bottom panel: The resulting change of the log-likehood expression for scans in the combinations of the experimental systematic uncertainties, being the most favorable for the models (indicated by a line connecting non-modification point with the plane in the middle panel). The figures come from \cite{PierreAuger:2024neu}.}
    \label{fig:RhadDeltaXmax}
  \end{figure}
  
\section{Inconsistencies between Models and Combined Measurements of Ground Signal and \Xmax}
In general, combining independent measurements of the longitudinal and lateral profiles provide the best tool how to test the validity of air-shower modelling using measured data.
The correlation of the ground signal and \Xmax is a quasi-model-independent estimator of how the primary beam is mixed.
It was shown in Ref.~\cite{MixedAnkle} that the \qgsii model provides too light mix of primary species (protons and helium nuclei only) when the fractions of primary species are fitted to the measured \Xmax distributions in the energy range $10^{18.5}-10^{19.0}\eV$.
The measured correlation factor between the ground signal and \Xmax is incompatible with such a mix of light primary particles.
The models \eposlhc and \sib{2.3d} predict deeper \Xmax scale and therefore the interpreted mass composition is heavier and more mixed than in case of the \qgsii model, being consistent with the observed correlation factor.

Taken the previous indications that not only the hadronic part of the ground signal ($S_\text{had}\rightarrow S_\text{had}\times\Rhad$), but also the predicted \Xmax scale might be incorrectly estimated ($\Xmax\rightarrow\Xmax$\,+\,$\DeltaXmax$) by the current models of hadronic interactions, we developed the following test.
We used mass-composition fits of five two-dimensional distributions of the ground signal and \Xmax with Monte-Carlo templates corresponding to five zenith-angle bins in the energy range $10^{18.5}-10^{19.0}\eV$ by minimizing the log-likelihood function $\ln\mathcal{L}$, for details see \cite{PierreAuger:2024neu}.

The results show that $>5\sigma$ improvement in the description of measured data is obtained by current models of hadronic interactions when additionally to the \Rhad also \DeltaXmax is considered.
The best-fit modification parameters are shown in the top panels of Fig.~\ref{fig:RhadDeltaXmax} with statistical contours and bands corresponding to the projected systematic uncertainties.
For all three current versions of models of hadronic interactions, the predicted hadronic part of the ground signal and also the \Xmax scale need to be increased to reliably describe the data of the Pierre Auger Observatory.
The deeper \Xmax scale has a consequence of a heavier mass composition needed to describe the data, which means less muons to be additionally generated than in the previous studies.
This alleviation of the muon problem to approximately its half (15-25\%) is also in line with the preliminary results of the shower universality method that provides estimate of the muon scale in models with a suppressed sensitivity to the \Xmax scale due to additional information from the fluorescence telescopes \cite{MaxRhadShowerUniv_ICRC2023}.

For the first time, the ability of the current models of hadronic interactions to describe measured data was disfavored by more than $5\sigma$.
Actually, for the most favorable combination of experimental systematic uncertainties (even out of the range of the quoted systematic uncertainties), none of the studied models is able to describe the measured data without a modification in the hadronic part of the predicted ground signal or \Xmax at the level of $5\sigma$, see the middle and bottom panels of Fig.~\ref{fig:RhadDeltaXmax}.

\section{Summary}
The Pierre Auger Observatory is currently the best air-shower instrument to study hadronic interactions above $\sqrt{s}\simeq 70$~TeV given the combination of surface and fluorescence detectors, recorded event statistics of UHECR and applied reconstruction methods.
Inconsistencies in the description of measured data by the current models of hadronic interactions have been observed at a level of $\leq3\sigma$ before, but only recently more than $5\sigma$ claim has been achieved between 3 and 10\EeV \cite{PierreAuger:2024neu}. 
The so-called muon problem has been found more complex.
The problem in the amount of generated muons has been alleviated to its half approximately for the cost of a new problem appearing in the predicted scale of \Xmax.

The deeper \Xmax scale predicted by the models, if proven at lower and higher energies than studied so far, would suggest much heavier mass composition than has been usually considered so far, as already indicated in Ref.~\cite{HeavyMetal_UHECR2024}.
The data of the upgraded Pierre Auger Observatory, AugerPrime \cite{AugerPrime}, will provide better discrimination of the electromagnetic and muon signal in the surface detectors, which will validate the assumptions of the method and probe such mass-composition scenarios at the higher energies.

\acknowledgments
The work was supported by the Czech Academy of Sciences: LQ100102401, Czech Science Foundation: 24-13049S, Ministry of Education, Youth and Sports, Czech Republic: LM2023032, CZ.02.1.01/0.0/0.0/16\_013/0001402, CZ.02.1.01/0.0/0.0/18\_046/0016010 and CZ.02.1.01/0.0/0.0/ 17\_049/0008422, FORTE CZ.02.01.01/00/22\_008/0004632. 

\bibliographystyle{JHEP}
\bibliography{bibtex.bib}

\end{document}